\documentclass{nature}

\usepackage{graphicx}
\usepackage{multirow}
\graphicspath{ {images/} }
\DeclareGraphicsExtensions{.png,.pdf,.eps}

\usepackage{epstopdf}
\usepackage{amsmath,amssymb}
\usepackage{hyperref}
\newcounter{firstbib}

\newcommand{\apj}{Astrophys. J.}
\newcommand{\icarus}{Icarus}
\newcommand{\apjl}{Astrophys. J.}

\newcommand{\aap}{Astron. Astrophys.}
\newcommand{\mnras}{Mon. Not. R. Astron. Soc.}
\newcommand{\aj}{Astron. J.}
\newcommand{\nat}{Nature}

\title{Atmospheric reconnaissance of the habitable-zone Earth-sized planets orbiting TRAPPIST-1}

\author{Julien de Wit$^{1,*}$, Hannah R. Wakeford$^{2,3,*}$, Nikole K. Lewis$^{3,*}$, Laetitia Delrez$^{4}$, Micha\"{e}l Gillon$^{5}$, Frank Selsis$^{6}$, J\'{e}r\'{e}my Leconte$^{6}$, Brice-Olivier Demory$^{7}$, Emeline Bolmont$^{8}$, Vincent Bourrier$^{9}$, Adam J. Burgasser$^{10}$, Simon Grimm$^{7}$, Emmanu\"{e}l Jehin$^{5}$,  Susan M. Lederer$^{11}$, James E. Owen$^{12}$,  Vlada Stamenkovi\'{c}$^{13,14}$ \& Amaury H. M. J. Triaud$^{15}$}

\begin{document}

\maketitle

\textsl{$^*$These authors contributed equally to this work.}
\begin{affiliations}
 \item Department of Earth, Atmospheric and Planetary Sciences, Massachusetts Institute of Technology, 77 Massachusetts Avenue, Cambridge, Massachusetts 02139, USA;
 \item Astrophysics Group, University of Exeter, Physics Building, Stocker Road, Devon, EX4 4QL UK;
 \item Space Telescope Science Institute, 3700 San Martin Drive, Baltimore, MD 21218, USA;
  \item Astrophysics Group, Cavendish Laboratory, 19 J J Thomson Avenue, Cambridge, CB3 0HE, UK;
 \item Space sciences, Technologies and Astrophysics Research (STAR) Institute, Universit\'{e} de Li\`{e}ge, All\'{e}e du 6 Ao\^{u}t 19C, 4000 Li\`{e}ge, Belgium;
  \item Laboratoire d'astrophysique de Bordeaux, Univ. Bordeaux, CNRS, B18N, all\'{e}e Geoffroy Saint-Hilaire, 33615 Pessac, France;
\item University of Bern, Center for Space and Habitability, Sidlerstrasse 5, CH-3012, Bern, Switzerland;
 \item Astrophysics Division of CEA de Saclay, Orme des Merisiers, Bat 709, 91191 Gif sur Yvette, France;
 \item Observatoire de l’Universit\'{e} de Gen\`{e}ve, 51 chemin des Maillettes, 1290 Sauverny, Switzerland;
 \item Center for Astrophysics and Space Science, University of California San Diego, La Jolla, California 92093, USA;
 \item NASA Johnson Space Center, 2101 NASA Parkway, Houston, Texas, 77058, USA;
 \item Astrophysics Group, Imperial College London, Blackett Laboratory, Prince Consort Road, London SW7 2AZ, UK;
 \item Jet Propulsion Laboratory, California Institute of Technology, Pasadena, CA 91109, USA;
 \item Division of Geological and Planetary Sciences, California Institute of Technology, Pasadena, CA 91125, USA;
 \item School of Physics \& Astronomy, University of Birmingham, Edgbaston, Birmingham B15 2TT, UK.
\end{affiliations}

\begin{abstract}
Seven temperate Earth-sized exoplanets readily amenable for atmospheric studies transit the nearby ultra-cool dwarf star TRAPPIST-1\cite{gillon2016, gillon2017}.  Their atmospheric regime is unknown and could range from extended primordial hydrogen-dominated to depleted atmospheres\cite{owen2013, jin2014, luger2015, owen2016}. Hydrogen in particular is a powerful greenhouse gas that may prevent the habitability of inner planets while enabling the habitability of outer ones\cite{Sagan1977,pierrehumbert2011,owen2016}. An atmosphere largely dominated by hydrogen, if cloud-free, should yield prominent spectroscopic signatures in the near-infrared detectable during transits. Observations of the innermost planets ruled out such signatures\cite{dewit2016b}. However, the outermost planets are more likely to have sustained such a Neptune-like atmosphere\cite{Bolmont2016,Bourrier2017}. Here, we report observations for the four planets within or near the system's ``Habitable Zone'' (HZ)--the circumstellar region where liquid water could exist on a planetary surface\cite{kasting1993,Zsom2013,kopparapu2016}. These planets do not exhibit prominent spectroscopic signatures at near-infrared wavelengths either, which rules out cloud-free hydrogen-dominated atmospheres for TRAPPIST-1 d, e and f with significance of \textbf{8, 6 and 4$\sigma$}, respectively. Such an atmosphere is instead not excluded for planet g. As high-altitude clouds and hazes are not expected in hydrogen-dominated atmospheres around planets with such insolation\cite{hu2014, morley2015}, these observations further support their terrestrial and potentially habitable nature.

\end{abstract}

We observed transits of TRAPPIST-1 planets d, e, f, and g with four visits of the Hubble Space Telescope (HST). Each of the visits contained two planetary transits (Figure 1), planets d and f in visits 1 (4 December 2016) and 3 (9 January 2017), and planets e and g in visits 2 (29 December 2016) and 4 (10 January 2017). The observations were conducted using the `forward' scanning mode with the near-infrared (1.1-1.7$\mu$m) G141 grism on the wide-field camera 3 (WFC3) instrument (see Methods). We capitalized on the frequency of the transit events in the TRAPPIST-1 system to select observation windows encompassing transits from two different planets, thereby optimizing the time allocation. The time sensitivity of these observations (TRAPPIST-1's visibility window closing in January 2017) combined with our ``multiple-transit-per-visit'' approach constrained us to perform exposures when HST crossed through the South-Atlantic Anomaly (SAA). 

Visits 1, 3, and 4 contain SAA crossing events which forces HST into GYRO mode, where its fine pointing ability is lost. The loss of fine pointing during and following the SAA crossing events cause the spectral position on the detector to change over time. In addition, during the SAA crossing a greater number of cosmic ray hits are introduced to the observations/exposures. We use the IMA output files from the CalWF3 pipeline and correct for this by cross-correlating each spectral read in the individual exposures and interpolating (see Methods). The raw light curves present primarily ramp-like systematics on the scale of HST orbit-induced instrumental settling discussed in previous WFC3 studies\cite{deming2013,Kreidberg2014,wakeford2016} (Figure 1). We chose here to not discard the first orbit of each visit--which is affected by larger systematics--but rather to develop a holistic systematic model allowing us to account for the time-dependent effects observed across the orbits of a visit\cite{wakeford2016,wakeford2017} to prevent reducing the observation baseline (see Methods). 

Despite the SAA crossing events in three of our visits, we mostly achieve per-orbit/visit precisions on par with what was achieved for the May 4th, 2016 observations of planets b and c. Summing over the entire WFC3 spectral range, we derived a ``white'' light curve across WFC3/G141's bandpass and reached an averaged SDNR (Standard Deviation of the Normalized Residuals) of 220 part per million (p.p.m.) over 21 of the 23 orbits (two orbits were heavily affected by the SAA crossing, see Methods), which is 1.5$\times$ the photon noise limit.  We reduced, corrected for instrumental systematics, and analyzed the data using independent methods presented in previous studies \cite{Gillon2012,wakeford2016,Demory2015,dewit2016b}. The independent analyses conducted by sub-groups of our team yield consistent results, which we report below.

We first analyzed the white light curves to measure transit depths and timings for comparison with previous observations. In retrieving these parameters, we treated each visit separately and simultaneously fitted for the two transits in each visit while accounting for instrumental systematics following Ref.\cite{dewit2016b}. Due to the reduced phase coverage of HST observations, we fixed the system's parameters to the values reported in the literature\cite{gillon2017} while estimating the transit times and depths. However, we let the band-integrated limb-darkening coefficients\cite{dewit2016b} (LDCs) and the orbital inclinations\cite{gillon2017} float under the control of priors, to propagate their uncertainties on the transit depth and time estimates with which they may be correlated. We report the transit depth and time of transit center estimates in Table\,\ref{tab:fit_param}. We note that the transit depth of planet d during visit 3 is poorly constrained and significantly affected by a 20-pixel long drift of the spectrum over the detector due to SAA crossing at the beginning of the orbit covering this transit. Although the effect of SAA crossing--and of the resulting GYRO mode--can be corrected if it occurs either during an orbit, or at its end, they cannot be corrected with high precision if the crossing occurs at the very beginning of an orbit. Beside this transit depth, all others are in agreement, within 2$\sigma$, with the values reported in the literature\cite{gillon2017}. We reach a precision of $\sim$40 sec on the transit timings for all but the transit of planet e during visit 4, for which neither the ingress or the egress of the planet's transit is recorded.

We then analyzed the light curves in 10 spectroscopic channels (1.15--1.65\,$\mu$m), fitting for wavelength-dependent transit depths, instrumental systematics, and stellar baseline levels. We use the wavelength-dependent priors on the LDCs reported in Ref.\cite{dewit2016b}. Our pipelines lead to an average SDNR of 520 p.p.m. per 112-second exposure (see Methods) on the spectrophotometric time series split in 10 channels (resolution = $\lambda/\Delta\lambda =\simeq 33$). We derived the transmission spectra of planets e and f jointly from visits 2 and 4 and visits 1 and 3, respectively. Due to increased scatter and cosmic ray hits during the SAA passes we were unable to derive a transmission spectrum from visit 3 for planet d or from visit 4 for planet g and used, thus only visit 1 and visit 2, respectively. The resulting transmission spectra are shown in Figure 2. 

The individual transmission spectra show no significant features. A comparison to aerosol-free versions of H$_2$-dominated atmospheres like those of the Solar System giant planets allow us to rule out such atmospheres at 8, 6, and 4 $\sigma$ for TRAPPIST-1 d, e, and f respectively. The current data results in only a 2-$\sigma$ confidence level for planet g, which is not significant enough to rule out this scenario for the planet. As for planets b and c\cite{dewit2016b}, many alternative atmospheric scenarios are consistent with the data such as H$_2$O-, N$_2$-, or CO$_2$-dominated atmospheres (shown respectively in blue, green, and grey in Figure 2), tenuous atmospheres composed of a variety of chemical species\cite{owen2013,jin2014,luger2015,owen2016,leconte2015}, or atmospheres dominated by aerosols\cite{morley2015}. The consistency of HST/WFC3's transit depth estimates with those of Spitzer/IRAC/4.5$\mu$m\cite{gillon2017} implies the lack of significant absorption features between the two different spectral ranges covered\cite{Sing2016}, thereby further suggesting the absence of clear H$_2$-dominated atmospheres. 

Hydrogen is a powerful greenhouse gas, and its presence in significant amounts in an atmosphere therefore affects a planet's habitability. The predominance of atmospheric hydrogen shapes the inner and outer edge of the Habitable Zone (HZ)\cite{pierrehumbert2011,owen2016}---the circumstellar region where water could stay liquid on a planetary surface\cite{kasting1993,Zsom2013,kopparapu2016}. While a significant amount of H$_2$ could prevent the atmospheres of TRAPPIST-1's outer planets to freeze, it would lead to high surface temperatures and pressures for the inner planets that are incompatible with liquid water. In order to be habitable, the latter must therefore have lost most of their atmospheric hydrogen or have never accreted/outgassed significant amounts of hydrogen in the first place\cite{owen2016}. 

Given the irradiation levels experienced by the Earth-sized planets in TRAPPIST-1's HZ, theory suggests that the probability to form aerosols in a hydrogen-rich atmosphere at the pressures probed by the transmission observations presented here is low\cite{hu2014, morley2015}. If exact, aerosol formation theories thus allow us to conclude that TRAPPIST-1~d, e, f, and g do not harbor a hydrogen-dominated atmosphere and are terrestrial and potentially habitable. 

The next milestone in the characterization of the TRAPPIST-1 planet atmospheres require spectroscopic measurements that enable the identification of aerosols--via, e.g., attenuation signatures in the planets' transmission spectra\cite{wakeford2015}--and atmospheres with larger mean molecular weights. This milestone will be possible with the next generation of observatories\cite{gillon2016,barstow2016,Morley2017}, notably the James Webb Space Telescope (JWST). 

As the exploration of ``habitable-zone'' and ``temperate'' exoplanet atmospheres is initiated over the next decade, a new light will be progressively shed on the concept of the ``habitable zone''. This important concept is currently poorly constrained because its dependence to key parameters such as the host star type and the planet orbital configuration (incl. tidal locking) have not been mapped with the current sample at hand--the Solar System planets. New perspectives from configurations vastly different from those found in the Solar System will therefore be pivotal to improve our understanding of a planet habitability and refine the concept of habitable zone\cite{Zsom2015}.

\newpage
\bibliographystyle{naturemag}

\begin{addendum}
 \item This work is based on observations made with the NASA/ESA Hubble Space Telescope that were obtained at the Space Telescope Science Institute, which is operated by the Association of Universities for Research in Astronomy, Inc. These observations are associated with program GO-14873 (principal investigator J.d.W.), support for which was provided by NASA through a grant from the Space Telescope Science Institute. H.R.W. acknowledges funding from the European Research Council (ERC) under the European Union’s Seventh Framework Programme (FP7/2007-2013) / ERC grant agreement no. 336792, and funding under the STScI Giacconi Fellowship. This work was partially conducted while on appointment to the NASA Postdoctoral Program at Goddard Space Flight Center, administered by USRA through a contract with NASA. L.D. acknowledges support from the Gruber Foundation Fellowship. E.J. and M.G. are Research Associates at the Belgian Fonds (National) de la Recherche Scientifique (F.R.S.-F.N.R.S.). The research leading to these results has received funding from the European Research Council under the FP/2007-2013 ERC Grant Agreement n° 336480, and from the ARC grant for Concerted Research Actions, financed by the Wallonia-Brussels Federation. B.-O.D. acknowledges support from the Swiss National Science Foundation (PP00P2-163967). This work was also partially supported by a grant from the Simons Foundation (PI Queloz, grant number 327127). This project has received funding from the European Research Council (ERC) under the European Union's Horizon 2020 research and innovation program (grant agreement No. 679030/WHIPLASH). V.B. acknowledges the financial support of the SNSF. We thank D. Taylor, K. Stevenson, N. Reid, and K. Sembach for their assistance in the planning, execution, and/or analysis of our observations. J.d.W., H.R.W., and N.K.L. thank also the Howards-Lewis Team and F. Dory for their support and insightful contributions during the data-processing phase of this work.
 
\item[Author Contributions] J.d.W. and N.K.L. led the management of the survey. J.d.W. planned the observations. J.d.W and H.R.W. led the data reduction and analysis with the support of N.K.L., L.D., M.G., and B.-O.D. J.d.W. led the data interpretation with the support of H.R.W., N.K.L., V.S., J.L., J.E.O., and F.S. Every author contributed to write the manuscript and/or the HST proposal behind these observations. 

\item[Competing Interests] The authors declare that they have no competing financial interests.

\item[Code Availability] Conversion of the UT times for the photometric measurements to the BJD$_{TBD}$ system was performed using the online program created by J. Eastman and distributed at\\ 
http://astroutils.astronomy.ohiostate.edu/time/utc2bjd.html. We have opted not to make available the codes used for the data extraction as they are currently a significant asset of the researchers' tool kits. However, subproducts of the data extractions such as corrected frames and extracted white and colored light-curves can be found on the MAST archive (archive.stsci.edu). We have opted not to make available all but one of the codes used for the data analyses for the same reason. The MCMC software used by M.G. to analyze independently the photometric data is a custom Fortran 90 code that can be obtained upon request.

\item[Data Availability Statement] The data that support the plots within this paper and other findings of this study are available on the Mikulski Archive for Space Telescopes (\href{https://mast.stsci.edu/portal/Mashup/Clients/Mast/Portal.html?searchQuery=trappist-1}{Link to TRAPPIST-1 observations}) and/or from the corresponding author upon reasonable request.

 \item[Correspondence] Correspondence and requests for materials
should be addressed to Julien de~Wit~(email: jdewit@mit.edu). 

\end{addendum}

\newpage

\begin{table*}
	\caption{Transit depths and timings for program HST-GO-14873.}
	\label{tab:fit_param}
		
	\centering\footnotesize{\begin{tabular}{c|c|cccc}
		\hline\hline
	&	& \textbf{TRAPPIST-1~d} & \textbf{TRAPPIST-1~e} & \textbf{TRAPPIST-1~f} & \textbf{TRAPPIST-1~g}\\
		
		\hline
\multirow{2}{*}{Visit 1} & Transit depth [p.p.m.] & 3984$\pm$87 & -- & 6227$\pm$192 & --\\
& Transit timing$\mathbf{^{\mathrm{a}}}$  & 726.84005$\pm$0.00041 & -- & 726.62108$\pm$0.00048 & --\\
\hline
\multirow{2}{*}{Visit 2} & Transit depth [p.p.m.] & -- & 4754$\pm$88 & -- & 7823$\pm$133\\
& Transit timing$\mathbf{^{\mathrm{a}}}$  & -- & 751.87016$\pm$0.00036 & -- & 751.83978$\pm$0.00047 \\
\hline
\multirow{2}{*}{Visit 3} & Transit depth [p.p.m.] & 8066$\pm$354$^{\mathrm{b}}$ & -- & 6452$\pm$172 & --\\
& Transit timing$\mathbf{^{\mathrm{a}}}$  & 763.28978$\pm$0.00055$^{\mathrm{b}}$ & -- & 763.44484$\pm$0.00049 & --\\
\hline
\multirow{2}{*}{Visit 4} & Transit depth [p.p.m.] & -- & 5005$\pm$101 & -- & 7739$\pm$219$^{\mathrm{b}}$\\
& Transit timing$\mathbf{^{\mathrm{a}}}$  & -- & 764.06713$\pm$0.00176 & -- & 764.19120$\pm$0.00061$^{\mathrm{b}}$ \\
		\end{tabular}}	 
\begin{list}{}{}
\item[$^{\mathrm{a}}$] {Timings are reported as barycentric Julian date (BJD)/barycentric coordinate time (TBD)-2,457,000.}
\item[$^{\mathrm{b}}$] {The transits of planets d and g during Visit 3 and Visit 4, respectively, were discarded due to strong systematics induced by a large drift of the stellar spectrum (see Methods).}
\end{list}


\end{table*}

\pagebreak
\clearpage
\newpage

\begin{figure}
\begin{center}
\vspace{-3cm}\hspace{-1cm}\includegraphics[width=15cm,height=!]{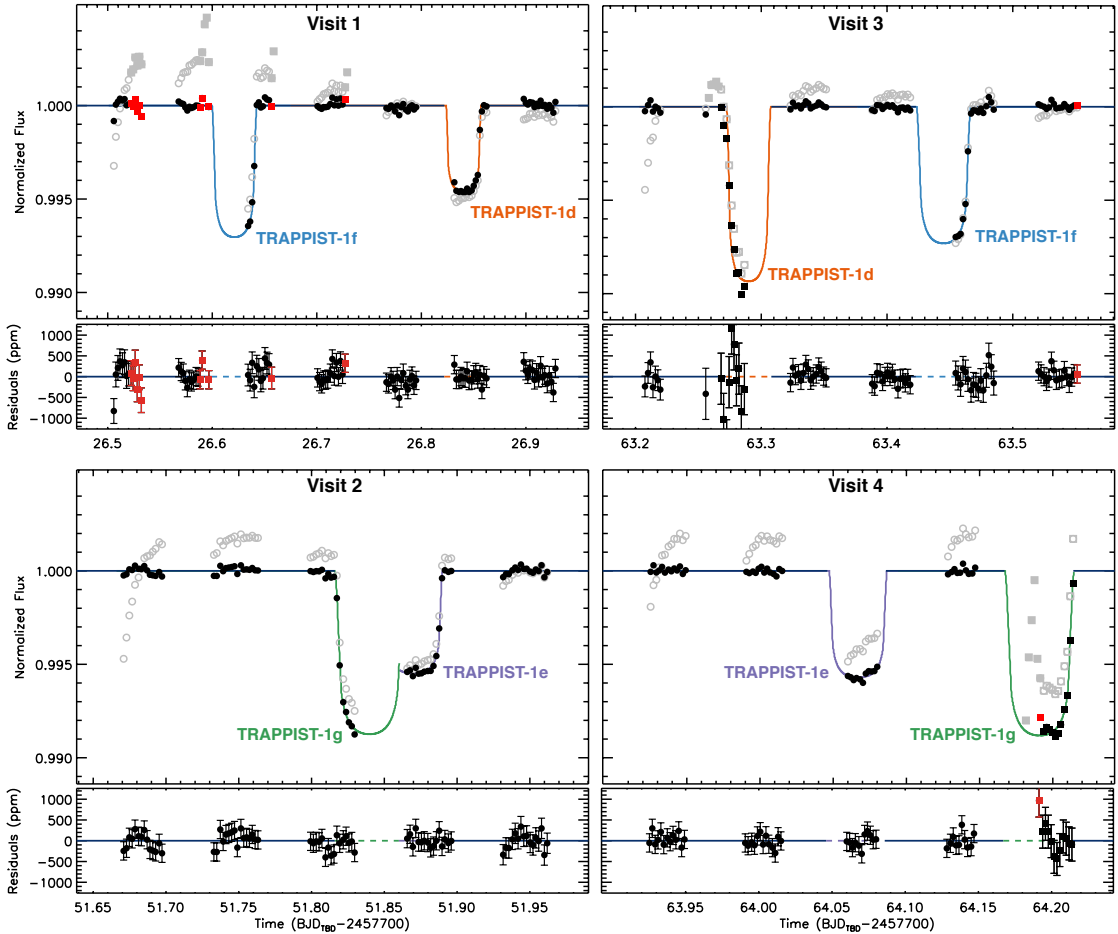}
\vspace{-1cm}\caption{Hubble/WFC3 white-light curves of the four  TRAPPIST-1 habitable-zone planets--TRAPPIST-1\,d,\,e,\,f, and \,g--over four visits. In the top panel of each visit the raw normalized light curves (gray) are shown with the systematic corrected light curves (black) against the best-fit transit model (colored solid line). In visits 1, 3, and 4, the observations were taken during the SAA crossing indicated by filled gray points in the raw data and red points in the corrected data. During and following the SAA crossing, HST enters GYRO mode, and we show each of these impacted exposures as squares in the datasets. The bottom panel of each visit shows the best fit residuals with their 1$\sigma$ error bars, we again indicate where the SAA and GYRO exposures occurred during each visit, and the dashed horizontal lines indicate where the transit occurred.}\label{fig:1}
\end{center}
\end{figure}

\begin{figure}
\hspace{-2cm}\includegraphics[width=20cm,height=!]{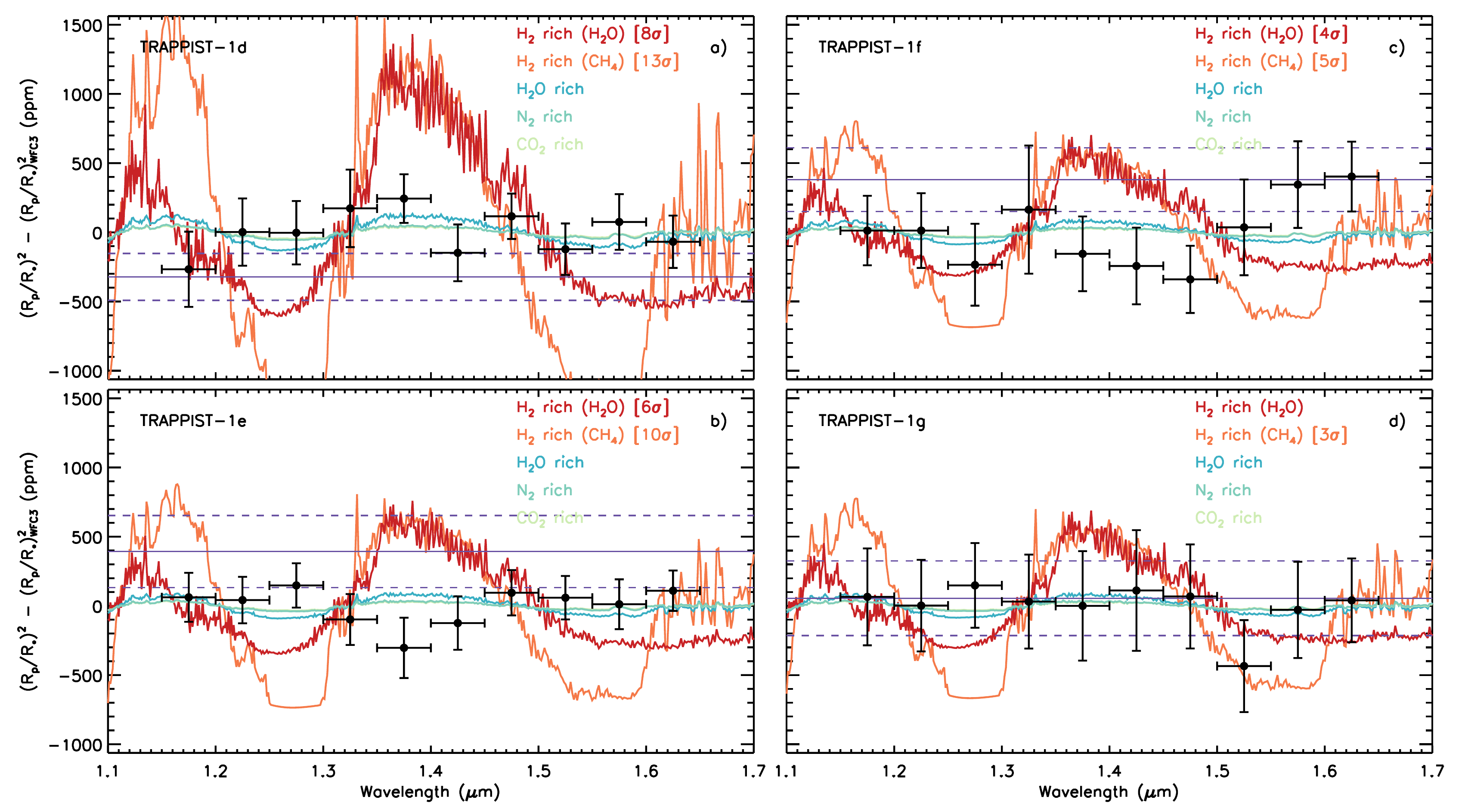}
\caption{ Transmission spectra of TRAPPIST-1\,d, e, f, and g compared to synthetic H$_2$-, H$_2$O-, CO$_2$-, and N$_2$-dominated atmospheres. HST/WFC3 measurements are shown as black circles with 1$\sigma$ errorbars. Each spectrum is shown shifted by its average over WFC3's band. The measurements are inconsistent with the presence of a cloud-free H$_2$-dominated atmosphere at greater than 3$\sigma$ confidence for planets d, e, and f (only the values larger than 3$\sigma$ are reported in the legends). The measurements for all four planets are consistent with the multiple scenarios of compact atmospheres explored as well as with the transit depths obtained with Spitzer/IRAC at 4.5$\mu$m\cite{gillon2017} (purple line with 1$\sigma$ errorbars).}  \label{fig:2}
\end{figure}

\pagebreak
\clearpage
\newpage

\begin{methods}

\subsection{HST WFC3 Observations.}
We observed the transits of TRAPPIST-1\,d, e, f, and g over the course of four visits--composed respectively of 7, 5, 6, and 5 orbits--each containing two of the planetary transits. Visits 1 and 3 contain the transits of planets d and f. Visits 2 and 4 contain the transits of planets e and g. Observations were conducted using HST WFC3 IR G141 grism (1.1--1.7$\mu$m) in forward spatial scanning mode\cite{mccullough2012}, which leads to less instrumental systematics and should be favored for faint targets\cite{Kreidberg2014,knutson2014,dewit2016b}. Scans in each visit were conducted at a rate of $\approx$0.232 pixels per second, with a final spatial scan covering $\approx$26 pixels (3.38'') in the cross-dispersion direction on the detector. 

Visits 1, 3, and 4 contain observations conducted as HST crossed the South Atlantic Anomaly (SAA). During the SAA, HST enters GYRO mode and loses its fine pointing ability. (We flagged manually each frame acquired in the SAA or in GYRO mode as the relevant keywords in the file headers were not updated for each exposure.) During GYRO mode, the stellar spectrum may drift significantly across the detector array between exposures and even during the course of a single spatial scan. This results in a slanted spectral trace across the scan and detector for these affected exposures (Supplementary Figure 1). The shift caused in each exposure can be measured by cross-correlating the 1D spectrum from each read of the exposure to a template spectrum (Supplementary Figure 2), which is used to calculate the relative shift in pixel position in the dispersion direction across the detector for each read in each exposure for each visit (Supplementary Figure 3). In visit 2 where no GYRO mode exposures were taken, no pixel level shifts were measured between the successive reads or exposures. During visit 1, HST entered the SAA four times part of the way into each of the first four orbits, which left HST in GYRO mode for the rest of these orbits. In visits 3 and 4, HST crossed the SAA once at the start of an orbit (orbit 2 of visit 3 and orbit 5 of visit 4), which results in GYRO mode being used for the whole orbit causing significant shifts in the spectral position over the course of the observation sequences. These orbits were not fully recoverable; avoiding the initiation of orbits coinciding with a planetary transit in the SAA is the only caveat of the observational strategy used here.

To correct for this shift in the spectrum we use a spline interpolation to realign each of the successive reads for each exposure to a template spectrum, which is made from the zeroth read of the final exposure. Each of the exposure reads are then aligned such that the pixel column can be summed to produce the total flux over the whole exposure in the related 4.6-nm bin. During the SAA crossing, the exposures are subject to many times more cosmic ray hits than an average observation (Supplementary Figure 1). During the extraction, we correct for cosmic ray hits in two ways, first spatially by using the surrounding pixels in each 2D read to median replace the cosmic ray hits, secondly after realigning the extracted spectra by using the time axis comparing each pixel in the surrounding exposures and removing the detected cosmic rays. Over the observations, an average of 0.12\% of the pixels are corrected for cosmics.


\subsection{An Inclusive Ramp Model.}

The detector ramp of HST/WFC3 is orbit-dependent\cite{wakeford2016,wakeford2017}. Standard procedures consist either of discarding the first orbit\cite{Berta2012,Kreidberg2014,wakeford2016,dewit2016b} or in fitting a different ramp for the first orbit\cite{Line2016}. We chose here to introduce a new ramp model to account for the time-dependence of the ramp observed across the orbits of a visit (Figure 1) and prevent systematically discarding the first orbit, which would otherwise reduce the observation baseline. (We note that such an approach was independently introduced during the review process of this manuscript in Ref.\,\cite{Zhou2017}.) As observed recently for WASP-101 HST/WFC3 observations\cite{wakeford2017}, the variability of the detector ramp from an orbit to the next is enhanced when the electron count per exposed pixel remains too low ($<30,000 e^-$) to stabilize the charge-trapping effect during the first orbit. Therefore, the number of free charge traps settles over multiple orbits to an equilibrium value driven by the charge-trapping rate and the charge-release rate. Because the probability of a photon-generated charge to be trapped is directly proportional to the number of free charge traps, both the time-scale and the amplitude of the ``detector ramp'' settle as an exponential whose own time-scale--which is here larger than the duration of an orbit (i.e., $\sim$ 45 min)--is proportional to the number of free traps. We therefore use the physically-relevant following ramp model:
\begin{equation}
\frac{F_{obs}(t)}{F(t)} = r_v(t) r_o(t),
\end{equation}
where $r_v(t)$ is the ramp induced by the progressive stabilization of the charge-trapping,
\begin{equation}
r_v(t)= (1 + a_1 e^{-\frac{t}{a_2}}),
\end{equation}
$r_o(t)$, is the ramp induced by the charge-trapping effect (i.e., the ``traditional ramp'')
\begin{equation}
r_o(t)= (1 + a_3 e^{-\frac{t-t_{o_i}}{a_4 r_v(t)}}),
\end{equation}
$F_{obs}(t)$ is the flux observed, $F(t)$ is the incoming flux, $t$ is the time from the first exposure of the visit, $t_{o_i}$ is the time of the fist exposure of the $i^{th}$ orbit, which contains the exposure obtained at time $t$, and $\lbrace a_1,a_2,a_3,a_4\rbrace$ are the model parameters. 

Our model requires the same number of parameters as does fitting for a different ramp for the first orbit, while being physically-relevant and accounting for the orbit-dependence of the detector ramp. We find that our model combined with a second order polynomial in time over each visit is strongly favored ($\Delta$BIC$\geq$-25). Supplementary Figure 4 shows the best fits for visit 1 using our ramp model and different traditional ramps for the first orbit and the subsequent ones.

\subsection{HST WFC3 spectroscopy.}

The spectroscopic light curves are created for ten 0.05-$\mu$m bins from 1.15--1.65\,$\mu$m (Supplementary Figure 5). Out of the 23 orbits obtained in this program, two were significantly affected by the SAA as they started during the crossing (implying a high level of cosmic ray hits) and continued in GYRO mode (implying a large spectral drift). Both of these orbits occurred during a planetary transit; orbit 2 of visit 3 during a transit of planet d and orbit 5 of visit 4 during a transit of planet g. This, therefore, prevented us from capitalizing on the repeated observations for both planets to reach a higher precision. As we observed hints of offsets in the absolute transit depths of the planets e and f, we extracted their transmission spectra jointly from, respectively, visits 2 and 4 and visits 1 and 3 while allowing for an absolute transit depth offest from one visit to the next. Performing joint analyses allows one to further disentangle the visit-independent planetary signal from the visit-dependent systematics. Furthermore, it is a statistically more adequate approach than combining spectral estimates derived from individual visits in the context of systematics (i.e., red noise). We used the wavelength-dependent LDCs and analysis procedure (least-squares minimization fitting implementation to marginalize across a grid of systematic models, followed by adaptative MCMC implementations to sample the parameter posterior probability distributions)  introduced in ref.\cite{dewit2016b}. Doing so, we achieve for visits 1 to 4 an average SDNR of 545 p.p.m., 526 p.p.m., 493 p.p.m., and 494 p.p.m, on average across all spectroscopic light curves in each visit, respectively.

\subsection{Uncertainty estimates on the flux measurements.}

We estimate the uncertainty on the flux measurements from the standard deviation of the residuals of each individual orbit while accounting for a potentially reduced value due to the small sample size related to each orbit (up to 17 measurements). To do so, we prevent the uncertainty estimates to be lower than the program average out of SAA crossing (220 and 520 p.p.m. for the white and colored light curves, respectively). This approach allows the routine to automatically account for the larger SDNR of the early-SAA-crossing orbits (see Figure 1).

\subsection{Atmospheric analysis.}
We compared the derived transmission spectra of  TRAPPIST-1\,d,\,e,\,f and g to synthetic spectra representative of hydrogen-dominated atmospheres (Figure 2), like those of the Solar System giant planets. We simulated the synthetic spectra following Ref.\cite{dewit2016b} using the model introduced in ref.\cite{deWit2013}. We used the atmospheric compositions of the ``mini-Neptune' ($\mu=2.6$) and ``Halley world' ($\mu=14.9$) scenarios introduced in ref. \cite{benneke2012} to simulate the hydrogen-dominated and water-dominated atmospheres, respectively. We used a 2$\%$ abundance for the trace gas for the H$_2$-dominated atmosphere with methane (similarly to the ``mini-Neptune'' case featuring water as trace gas). We used an atmosphere with 80\% CO$_2$, 13\% CH$_4$, 3\% N$_2$, 2\% H$_2$O, and 2\% H$_2$ for the CO$_2$-rich atmosphere. We used for each planet the same mixing ratios and assume conservatively isocompositional and isothermal atmospheres in hydrostatic equilibrium. We used temperatures equal to the planets' equilibrium temperature assuming a 0.3 bond albedo (265\,K, 235\,K, 200\,K, and 180\,K for planets\,d,\,e,\,f and \,g, respectively). 

For the planetary masses, we conservatively use the maximum masses that would allow each of the planets to harbor hydrogen-dominated atmospheres (i.e., H$_2$-He envelopes greater than 0.01\% of their total masses given their radii\cite{howe2014}). These masses are referenced as ``conservative'' as they minimize the atmospheric scale height, the amplitude atmospheric signal in transmission, and thus also the significance level to which the synthetic scenarios can be ruled out by the measured spectra. These theoretical upper limits correspond to 0.4, 0.8, 1, and 1.15\,M$_{\oplus}$ for planets d to g, respectively. With this theory-based approach, our conclusion are thus also independent from the current mass estimates, which once refined could be used to derive quantitative atmospheric constraints.

\end{methods}

\newpage

\bibliographystyle{naturemag}

\newpage

\begin{figure}
\hspace{-3cm}\includegraphics[trim=20mm 20mm 20mm 20mm, clip=true,width=22cm,height=!]{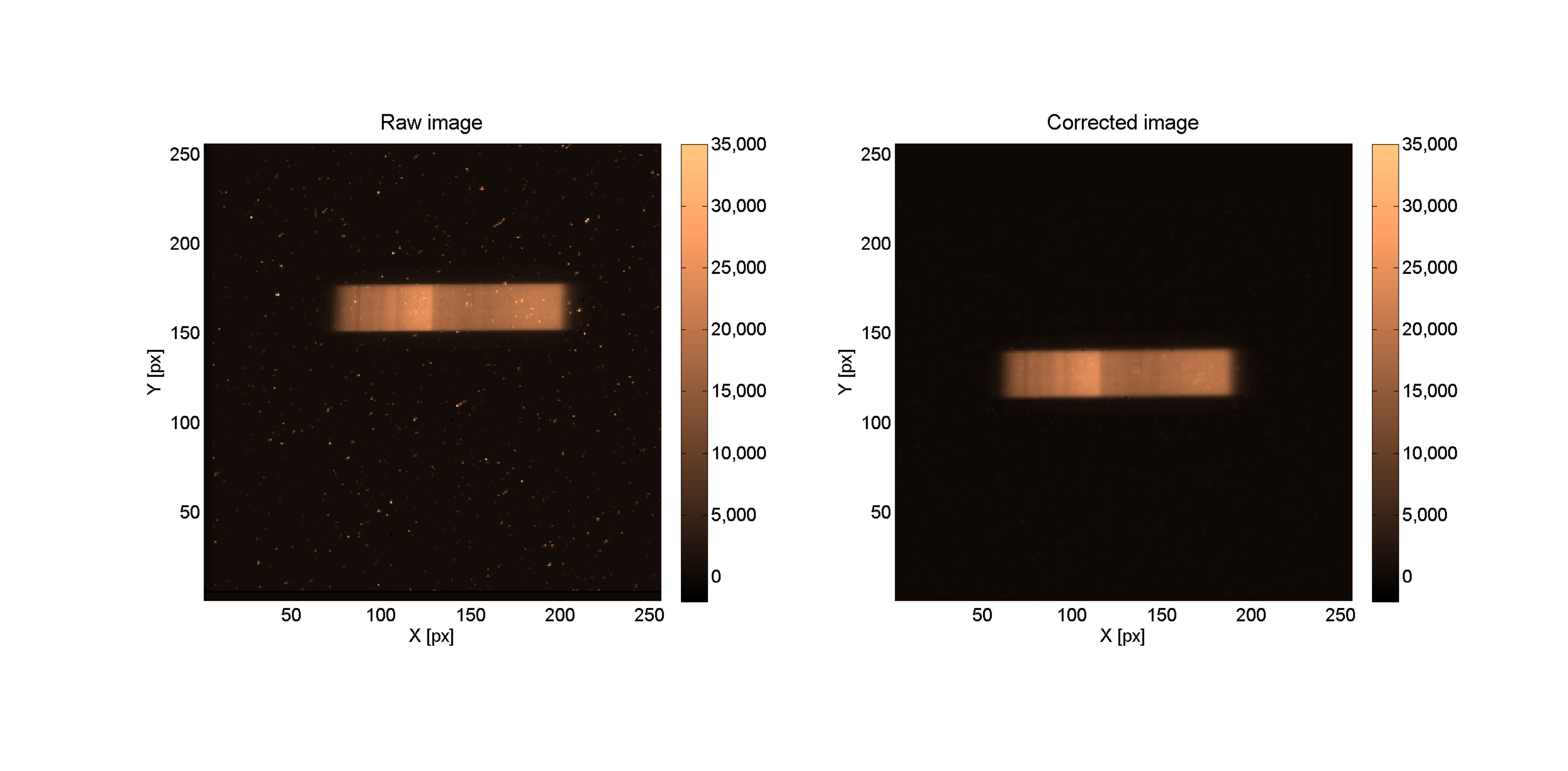}
\caption{Effect of SAA crossing on HST/WFC3 exposures. Left: Raw IMA frame corresponding to the exposure 56 of visit 4 (e56V4) taken during SAA crossing in GYRO mode, showing the increased cosmic ray interference and $\sim$2 pixel positional shift in the dispersion direction from the bottom to the top of the scan. The frame exhibits a high level of cosmic ray hits and a slanted spectral trace. Right: Corrected IMA e56V4 frame.}  \label{extendedfig:2}
\end{figure}

\begin{figure}
\hspace{-2cm}\includegraphics[width=18cm,height=!]{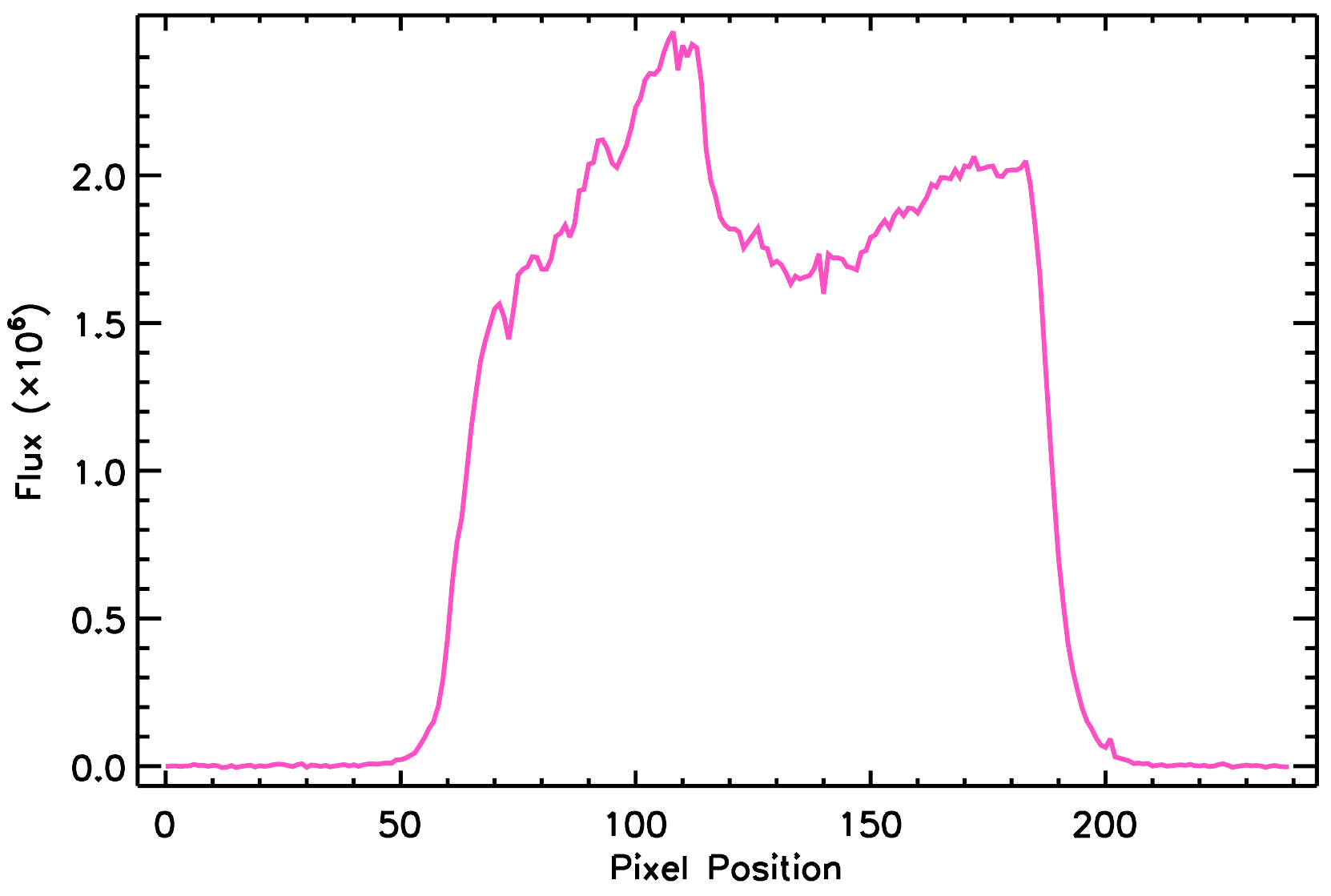}
\caption{Near-infrared  spectrum of TRAPPIST-1 as observed with HST/WFC3/G141. The flux unit is electron count per exposure and the spread along the x axis corresponds to 4.6 nanometers per pixel.}  \label{extendedfig:3}
\end{figure}

\begin{figure}
\hspace{-2cm}\includegraphics[width=18cm,height=!]{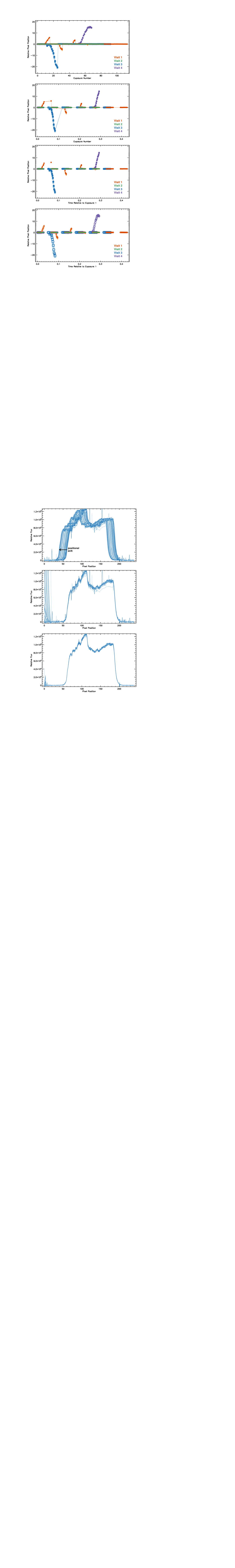}
\caption{Motion of the spectral trace in the dispersion direction on the detector. Pixel position of the spectrum for each read of each exposure (light to dark circles) in each visit, relative to a template spectrum made from the first read of the last exposure for each visit. The large pixel shifts show where HST entered GYRO mode prompted by an SAA crossing. The gaps are caused by the occultation of TRAPPIST-1 by the Earth. Visit 1 (orange) entered GYRO mode 4 times over the course of the observations. Visit 2 (green) did not enter GYRO mode. Visits 3 (blue) and 4 (purple) entered GYRO mode once during the visit over the course of an entire orbit, each coinciding with a planetary transit.}  \label{extendedfig:1}
\end{figure}

\begin{figure}
\hspace{-2cm}\includegraphics[width=18cm,height=!]{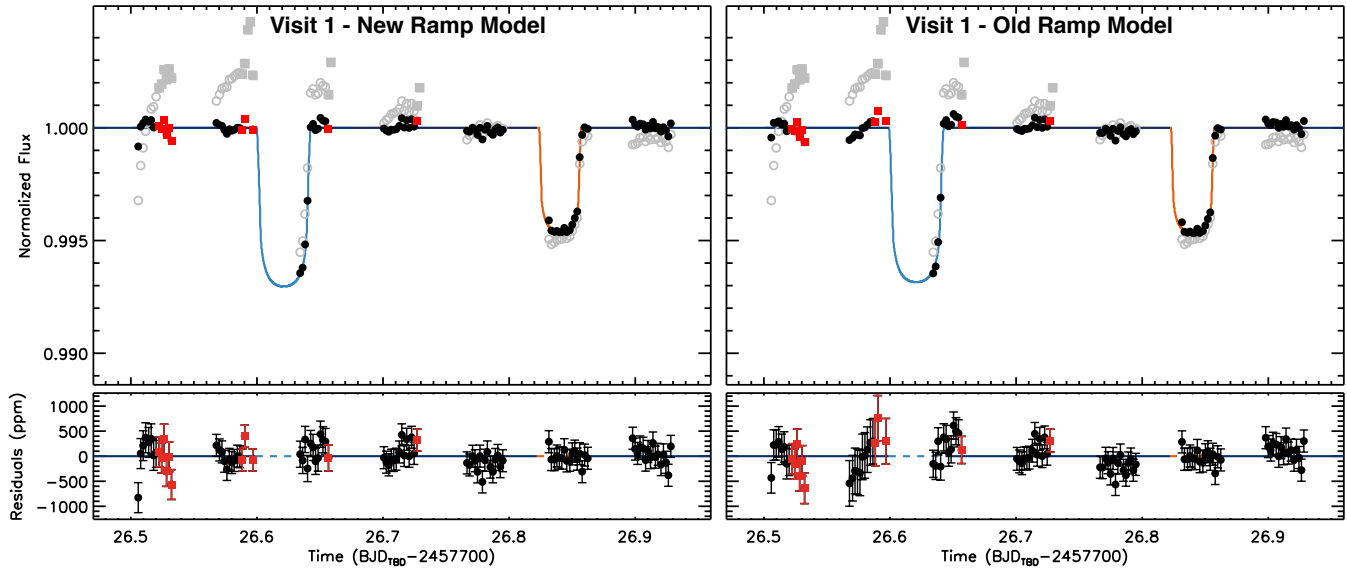}
\caption{A new ramp model for HST/WFC3. Hubble/WFC3 white light curves of visit 1 corrected using our new time-dependent ramp model (left) and the standard time-independent ramp model (right). The quantities reported are the same as in Figure 1. The inability to correct for the ramps of orbits 2-7 using a unique ramp--as is typically performed--is revealed by the progressively decreasing slope in the residuals of orbits 2 and 3 in the bottom right panel. Our model can correct for the amplitude and time-scale decrease of the ramp thereby correcting the ramp effect seen across all the orbits of a visit with a unique and inclusive model.} \label{extendedfig:4}
\end{figure}

\begin{figure}
\hspace{-2cm}\includegraphics[width=18cm,height=!]{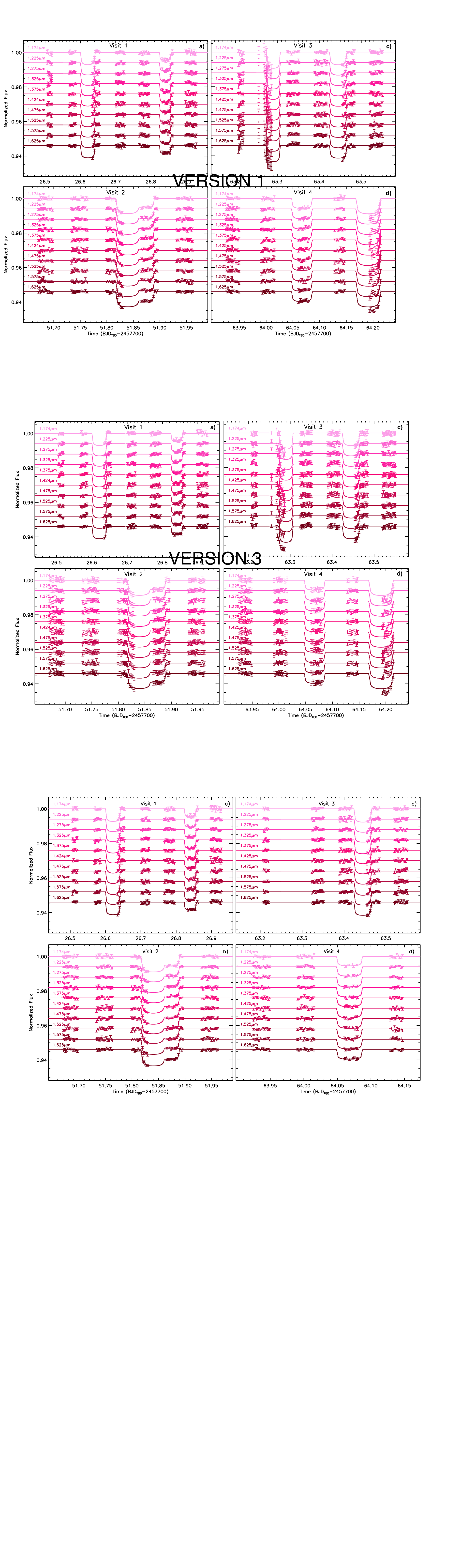}
\caption{ Hubble/WFC3 spectrophotometry of the four  TRAPPIST-1 habitable-zone planets--TRAPPIST-1\,d,\,e,\,f, and \,g--over four visits. Normalized and systematics-corrected data (points with their 1$\sigma$ error bars) and best-fit transit model (solid line) in 10 spectroscopic channels spread across the WFC3 band, offset for clarity (channel-averaged SDNR = $\sim$520 p.p.m.). The discarded measurements (notably the second orbit of visit 3 and the fifth orbit of visit 4) are not shown for clarity.}  \label{extendedfig:5}
\end{figure}




\end{document}